# Change of the charge modulation during superconducting transition in SmFeAsO$_{0.91}$F$_{0.09}$ seen by $^{57}$Fe Mössbauer spectroscopy

A. K. Jasek[1], K. Komędera[1], A. Błachowski[1], K. Ruebenbauer[1*], H. Lochmajer[2], N. D. Zhigadlo[3], and K. Rogacki[2,4]

[1]Mössbauer Spectroscopy Laboratory, Pedagogical University,
PL-30-084 Kraków, ul. Podchorążych 2, Poland

[2]Institute of Low Temperature and Structure Research, Polish Academy of Sciences,
PL-50-422 Wrocław, ul. Okólna 2, Poland

[3]Laboratory for Solid State Physics, ETH Zurich,
CH-8093 Zurich, Otto Stern Weg 1, Switzerland

[4]International Laboratory of High Magnetic Fields and Low Temperatures,
PL-53-421 Wrocław, ul. Gajowicka 95, Poland

[*]Corresponding author: sfrueben@cyf-kr.edu.pl

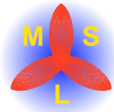

**Abstract**

Iron-based superconductor SmFeAsO$_{0.91}$F$_{0.09}$ has been investigated by the $^{57}$Fe Mössbauer spectroscopy versus temperature with the special attention paid to the region of the superconducting transition at about 47 K. Modulation of the electron charge density was found. It leads to the development of the charge density wave (CDW) and electric field gradient wave (EFGW). The modulation of CDW is enhanced in the temperature region of the superconducting gap opening, while the amplitude of EFGW is partly suppressed within this temperature region. This effect is exactly opposite to the similar effect in Ba$_{0.6}$K$_{0.4}$Fe$_2$As$_2$ superconductor. Hence, it seems that d electrons contribute significantly to the Cooper pair formation in both compounds as EFGW is perturbed within the temperature region of the superconducting gap formation.

## 1. Introduction

The compound SmFeAsO is a parent compound of the iron-based superconductors belonging to the '1111' family. Actually, the highest superconducting transition temperature of 56 K within iron-based superconductors is encountered for this family of compounds [1, 2]. Superconductivity is due to the presence of the corrugated Fe-As layers ordered in stacks without mutual inversion. Samarium and oxygen provide spacing between Fe-As layers, and the spacing is close to the optimal in order to develop high superconducting transition temperatures [2]. This metallic system is found as paramagnetic at high temperature due to the small itinerant magnetic moment generated by iron and localized magnetic moments of samarium with significant orbital component [3]. A high temperature phase is tetragonal at ambient pressure with the Fe-As layers oriented perpendicular to the tetragonal axis [4]. Upon cooling one observes some orthorhombic distortion at about 150 K followed by the development of the anti-ferromagnetic order of the itinerant magnetic iron moments at 144 K [2]. The latter moments develop spin density wave (SDW) incommensurate with the corresponding lattice period. This is a longitudinal SDW propagating perpendicular to the tetragonal (now orthorhombic) axis. Samarium orders anti-ferromagnetically at 5.6 K and some magnetic moment reorientation occur at temperature of 2.7 K [2].

Replacement of the oxygen by fluorine suppresses iron magnetic moment. Hence, the magnetic ordering temperature due to the itinerant electrons is lowered with the subsequent lowering of the transition to the orthorhombic phase as these two transitions are coupled by the magneto-elastic forces. Replacement of the oxygen increases conduction electron density as the divalent anion (oxygen) is replaced by the monovalent anion (fluorine) [5]. Hence, pockets of the Fermi surface are filled and SDW disappears leading to the diamagnetic contribution by the itinerant system. Magnetism of the 3d electrons and orthorhombic distortion disappear for about 4.5 at.% of oxygen replaced by fluorine [2]. Instead superconducting transition is observed with the transition temperature rising with the increasing concentration of fluorine. Magnetic ordering of samarium is weakly affected by mentioned replacement and one observes small lowering of the ordering temperature with increased fluorine content. Samarium is ordered in the superconducting state at sufficiently low temperature, while there is no 3d magnetic moment within superconductor [6]. Coexistence of the superconductivity and localized 4f magnetic moments order (even ferromagnetic) is well documented nowadays, albeit mainly for the pure spin states without significant orbital contribution present here [7, 8].

It was observed previously by $^{57}$Fe Mössbauer spectroscopy that some electronic charge modulation occurs in the iron-based superconductors like $Ba_{0.6}K_{0.4}Fe_2As_2$ belonging to the '122' family [9]. This modulation is sensitive to the superconducting transition and it is seen as s electrons charge density wave (CDW) and non-s electron (mainly d electrons) modulation seen as the electric field gradient wave (EFGW). Hence, it is interesting to look for similar effects in other family of the iron-based superconductors like '1111' family represented by $SmFeAsO_{0.91}F_{0.09}$.

## 2. Experimental

$SmFeAsO_{0.91}F_{0.09}$ was grown by solid state reaction starting from SmAs, FeAs, $SmF_3$, $Fe_2O_3$ and Fe by means of a high-pressure synthesis. The pulverized starting materials were sealed in a BN crucible, brought to a pressure of about 30 kbar at room temperature, heated within 1 h up to 1350 – 1450 °C, kept for 4.5 h at this temperature and finally quenched back to room temperature. Details about sample preparation and method used are given in Refs. [10, 11]. Resistivity and magnetization were studied as a function of temperature to reveal the superconducting properties of our sample. The results are shown in Figure 1. A transition to the superconducting state was observed at $T_{sc} = 47$ K, both for resistivity and magnetization, where the criteria used were the temperature at $0.5\rho_n$ ($\rho_n$ is the resistivity of the normal state value) and the temperature at the beginning of the diamagnetic signal, respectively. The calculated volume susceptibility was about -1 (+/- 0.1), for the sample density 7.4 g/cm$^3$ and the demagnetizing factor 0.1, which proves bulk superconductivity in our sample.

Absorber for the Mössbauer measurements was prepared in the powder form mixing 44 mg of the SmFeAsO$_{0.91}$F$_{0.09}$ with B$_4$C carrier. The thickness of the absorber amounted to 22 mg/cm$^2$ of SmFeAsO$_{0.91}$F$_{0.09}$. Commercial $^{57}$Co(Rh) was kept at room temperature, while the absorber was cooled by using Janis Research Co. SVT-400 cryostat with the long time temperature stability better than 0.01 K. Spectra were collected with the help of the MsAa-3 spectrometer equipped with the Kr-filled proportional counter. The He-Ne laser-based Michelson interferometer was used to calibrate the velocity scale. Measurement geometry, count-rate and single channel analyzer window were kept constant during uninterrupted series of measurements since 4.2 K till 72 K. Additional spectra were collected at 80 K and 300 K. Spectra were evaluated applying Mosgraf-2009 software suite [12]. Data were fitted within the transmission integral approach using the same model as in the case of Ba$_{0.6}$K$_{0.4}$Fe$_2$As$_2$ [9].

### 3. Results and discussion

Figure 1 shows resistivity of the SmFeAsO$_{0.91}$F$_{0.09}$ sample versus temperature. Rather typical metallic behavior is observed above 55 K and a rather broad transition to the superconducting state appears below this temperature with the onset at about 52 K. The transition is completed at 40 K and the zero resistant state is present down to 4 K. These properties clearly show that the sample is slightly underdoped, since the optimally doped SmFeAs(O,F) has transition onset at about 55 K. The magnetic susceptibility (see inset of Figure 1) obtained in the applied field of 5 Oe reveals a diamagnetic signal which begins at 47 K and develops rapidly below 40 K. The values of susceptibility observed at low temperatures prove the bulk superconductivity of our samples, as explained in Experimental. Note that samarium moments are invisible in the superconducting state as the applied field is below the first critical field for this superconductor.

Figure 2 shows selected $^{57}$Fe Mössbauer spectra versus temperature. Due to the fact that Fe-As sheets are widely separated one from another the electric quadrupole interaction is almost invisible and one observes somewhat broadened spectral singlet at high temperature with the total shift versus room temperature α-Fe being typical for the formally divalent iron in the metallic environment. A broadening is due to the charge density wave (CDW) incommensurate with the respective lattice period lying in the plane of the Fe-As sheet. A broadening has two components. Namely, the broadening due to the scatter of the electron density on the iron nuclei (formally s-electrons) seen via distribution of the isomer shift, and broadening due to the scatter of the spurious electric field gradient (EFG) seen via distribution of the electric quadrupole splitting. The latter effect is due to the scatter of the non-s electrons (mainly d-electrons) distribution in the vicinity of the iron nuclei and results in the electric field gradient wave (EFGW) [9]. The combined effect somewhat diminishes with lowering of the temperature probably due to the ordering of the light interstitials like oxygen and fluorine. Some magnetic components appear below about 28 K leading to the spectrum broadening. About 22 % of the cross-section area has some small average magnetic hyperfine field on iron of the order of 2.1 T at 20 K, while remaining spectral singlet is broadened as it partly contains some magnetic component with a very small average hyperfine field. The magnetic component has almost the same field as previously at 4.2 K, but it makes 36 % contribution to the cross-section. Remaining singlet is broadened probably due to the transferred field from already magnetically ordered samarium [6, 7]. Some fluctuation of the fluorine concentration across the sample is unavoidable, and therefore one sees some SDW even deeply in the superconducting part of the phase diagram, as some regions are superconducting and free of 3d magnetic moments, while other exhibit SDW without superconductivity [13, 14].

Figure 3 shows spectra within a region of the superconducting transition. The highest temperature 58 K is above onset of the transition, while the lowest temperature 28 K is below transition, but still above any magnetic ordering temperature. One can see a change of the spectra shape between 50 K and 48 K, i.e., just at the superconducting gap opening. A recovery (partial) is observed between 42 K and 38 K, where the gap reaches full development. The situation is similar to the situation observed for the '122' superconductor $Ba_{0.6}K_{0.4}Fe_2As_2$ [9].

Essential parameters of the resonant cross-section are gathered in Figure 4 as functions of the temperature – mainly in the region of the superconducting transition. The onset of the superconducting transition is marked by dashed vertical line at 47 K. The average total shift $S$ (versus room temperature α-Fe) behaves normally showing only typical second order Doppler shift (SOD) dependence on the temperature. The absorber linewidth Γ diminishes from the room temperature till about 58 K due to the increased order and jumps to the higher value across onset of the transition partly recovering upon completion of the transition. Hence, the modulation of the s-like CDW is enhanced across transition. Actually, the opposite effect was observed for $Ba_{0.6}K_{0.4}Fe_2As_2$, but of the comparable size [9]. A behavior of the dimensionless absorber thickness $t_A$ confirms above finding as the absorber linewidth and absorber thickness is inversely correlated each other. A dispersion of the spectral shift is insufficient to describe spectral shape (even in the non-magnetic regions) as one has components offset by the larger velocity span than plausible due to the isomer shift scatter. Hence, some EFGW is present like for $Ba_{0.6}K_{0.4}Fe_2As_2$. The EFGW shape was approximated the same way as for $Ba_{0.6}K_{0.4}Fe_2As_2$ except for the constant component $\varepsilon_0$ being practically absent here. Parameters describing EFGW shape $A$ and $\beta$ [9] are shown versus temperature in Figure 4 as well.

Figure 5 shows reduced absorber recoilless fraction $f/f_0$ (normalized to the recoilless fraction $f_0$ at 28 K) versus temperature in the superconducting transition region. The onset of the transition is marked by vertical line. The recoilless fraction remains constant across transition. Hence, lattice dynamics seems unaffected by a transition to the superconducting state. It is interesting to note that $f/f_0$ at 300 K is the same as for $Ba_{0.6}K_{0.4}Fe_2As_2$ [9]. This is a strong indication that iron dynamics is practically the same in both compounds as it relies on the "inner" dynamics of the almost the same Fe-As sheet. On the other hand, a dispersion of the electron density on the iron nuclei Δρ due to the CDW (s-like electrons dispersion) drops from high temperature to the low temperature due to the increased order and shows a hump across transition with partial recovery once the gap is fully developed. This is again an opposite effect to the one observed for $Ba_{0.6}K_{0.4}Fe_2As_2$ [9].

Figure 6 shows shape of the EFGW $AF_{max}^{-1}F(\mathbf{q}\bullet\mathbf{r})$ versus phase angle $\mathbf{q}\bullet\mathbf{r}$ in the plane of the Fe-As sheet for selected temperatures together with corresponding distributions $w(\varepsilon-\varepsilon_0)$ of the quadrupole shift $\varepsilon - \varepsilon_0$ with the assumption that $\varepsilon_0 = 0$ [9]. One can see that screening of the distant non-spherical charge increases with the lowering of the temperature, partly vanishes at the gap opening and recovers once the Bose condensate is separated from the remainder of the system. For $Ba_{0.6}K_{0.4}Fe_2As_2$ one has enhancement of the screening at the gap opening and recovery to the previous state once the Bose system is separated. It means that bosons are made of the non-s states and the latter states are oriented almost orthogonally each other for the systems studied, i.e., for $Ba_{0.6}K_{0.4}Fe_2As_2$ and $SmFeAsO_{0.91}F_{0.09}$, respectively. One can roughly conclude that d states play important role in the Cooper pairs formation for the iron-based superconductors studied. However, it seems that the coupling forces are still provided by the phonon field excitations.

## 4. Conclusions

The Mössbauer spectroscopy is sensitive to the superconducting transition in the iron-based superconductors via change of the electron charge density modulation. The incommensurate with the lattice period charge modulation is seen via dispersion of the isomer shift and via distribution of the electric field gradient. The first effect is caused by the s electrons and is often called CDW effect [15, 16]. The second effect is due to the non-s electrons (mainly d electrons) and is called EFGW [9]. On the other hand, spectral parameters dependent on the lattice dynamics like recoilless fraction and SOD are insensitive to the transition, as the lattice dynamics remains practically unchanged across superconducting transition.

One can observe narrowing of the CDW in the transition region with simultaneous broadening of EFGW (the case of $Ba_{0.6}K_{0.4}Fe_2As_2$) or the opposite effect, i.e., broadening of CDW with narrowing of EFGW (the case of $SmFeAsO_{0.91}F_{0.09}$). The basic difference between these two compounds is that $Ba_{0.6}K_{0.4}Fe_2As_2$ is obtained from the parent compound by hole doping (replacement of divalent cation by monovalent cation), while $SmFeAsO_{0.91}F_{0.09}$ is obtained by the electron doping (replacement of divalent anion by monovalent anion). Hence, the Fermi surface moves opposite way for above two cases.

## Acknowledgments

This work was supported by the National Science Center of Poland, DEC-2011/03/B/ST3/00446.

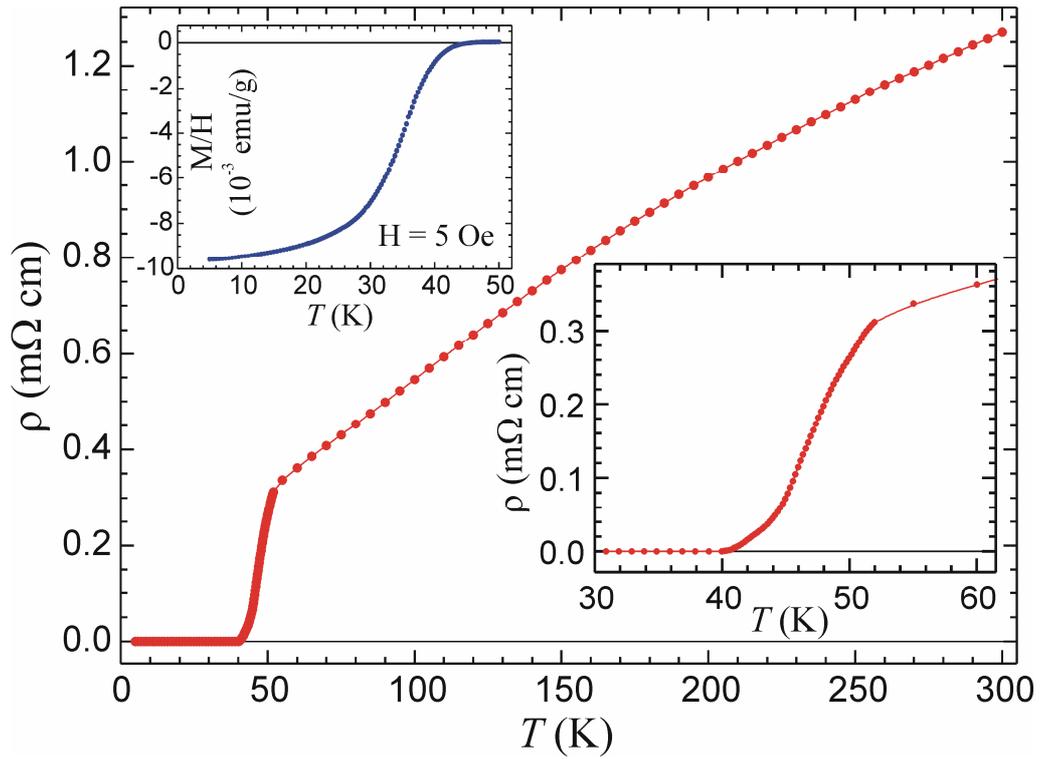

**Figure 1** Resistivity versus temperature for SmFeAsO$_{0.91}$F$_{0.09}$ sample. Right inset: resistivity in the vicinity of the transition to the superconducting state. Left inset: magnetic susceptibility versus temperature at the transition and in the superconducting state.

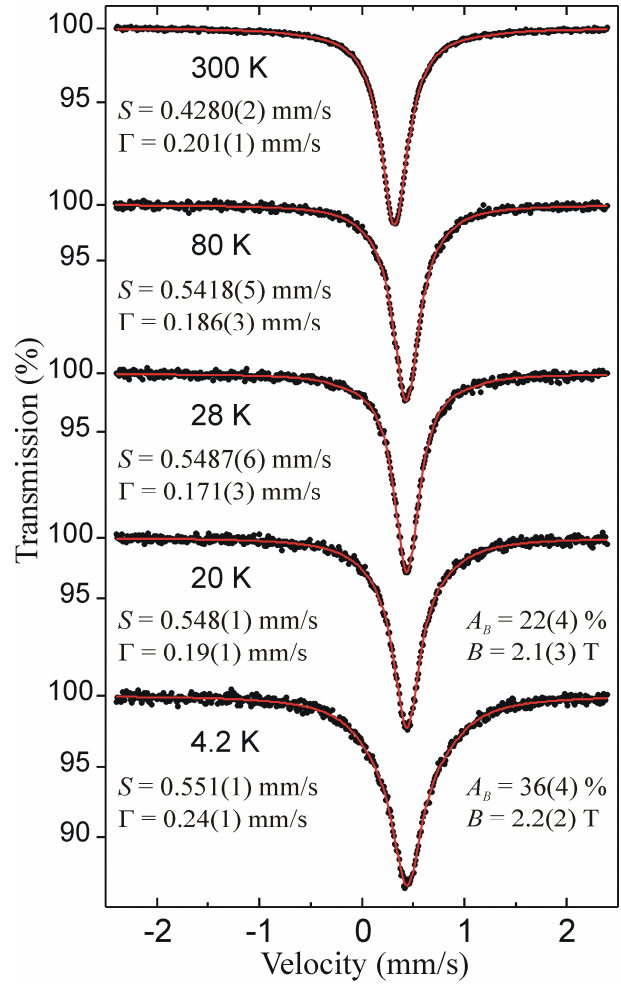

**Figure 2** Selected Mössbauer spectra of SmFeAsO$_{0.91}$F$_{0.09}$ obtained at various temperatures. The symbol $S$ stands for the spectral shift versus room temperature α-Fe, the symbol Γ denotes absorber linewidth, while $A_B$ stands for the contribution to the resonant cross-section due to the magnetically split component with the average hyperfine field $B$.

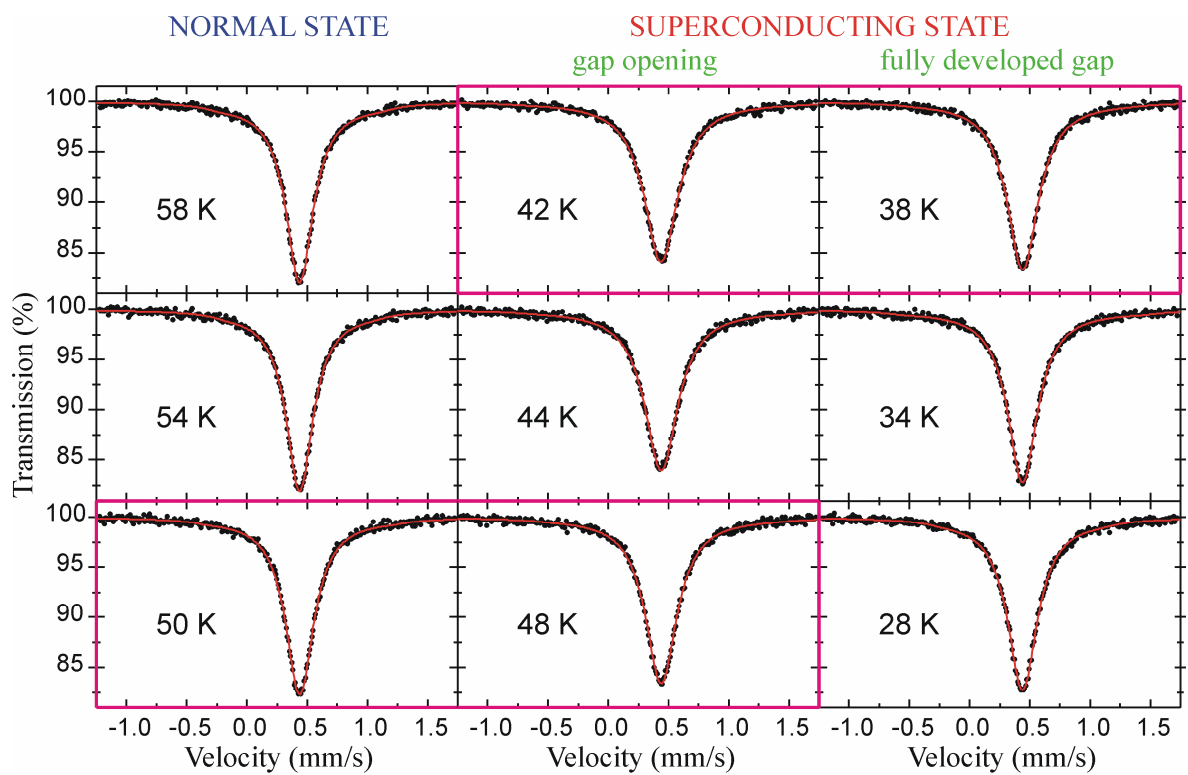

**Figure 3** Mössbauer spectra of SmFeAsO$_{0.91}$F$_{0.09}$ ($T_{sc} \approx$ 47 K) obtained across transition from the superconducting state to the normal state. Note the abrupt changes in the regions 38 K – 42 K and 48 K – 50 K.

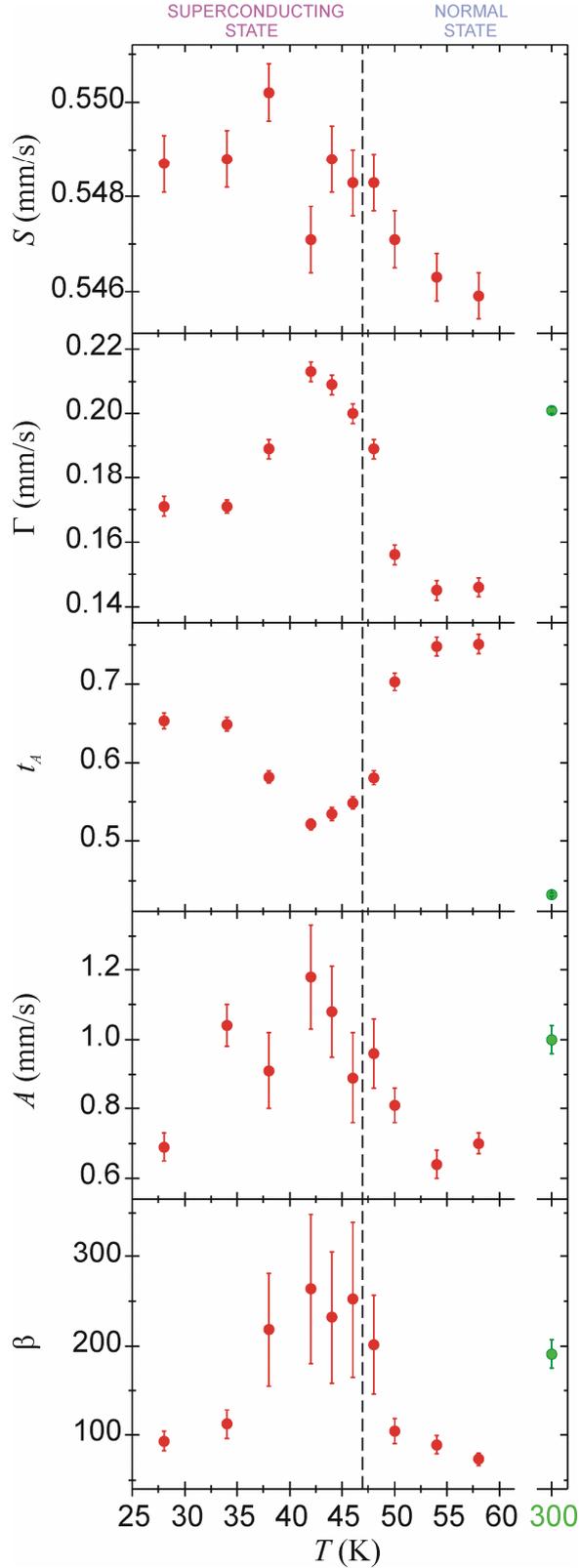

**Figure 4** Essential parameters derived from the Mössbauer spectra of SmFeAsO$_{0.91}$F$_{0.09}$ plotted versus temperature, i.e., spectral shift $S$ versus room temperature α-Fe, absorber linewidth $\Gamma$, dimensionless absorber thickness $t_A$, and two parameters describing shape of EFGW – $A$ and $\beta$. The spectral shift at 300 K amounts to +0.4280(2) mm/s. Dashed vertical line at 47 K marks onset of the superconducting transition.

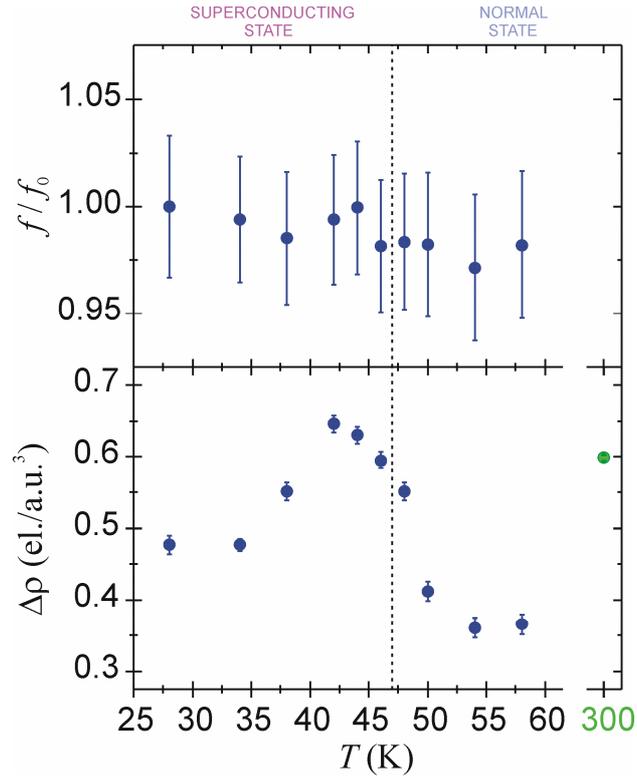

**Figure 5** Upper part shows relative recoilless fraction $f/f_0$ (normalized to the recoilless fraction $f_0$ at 28 K) plotted versus temperature. One obtains $f/f_0$ equal to 0.78(2) at 300 K. The lower part shows scatter of the electron density (dispersion of CDW) $\Delta\rho$ on the iron nuclei. Dashed vertical line at 47 K marks onset of the superconducting transition.

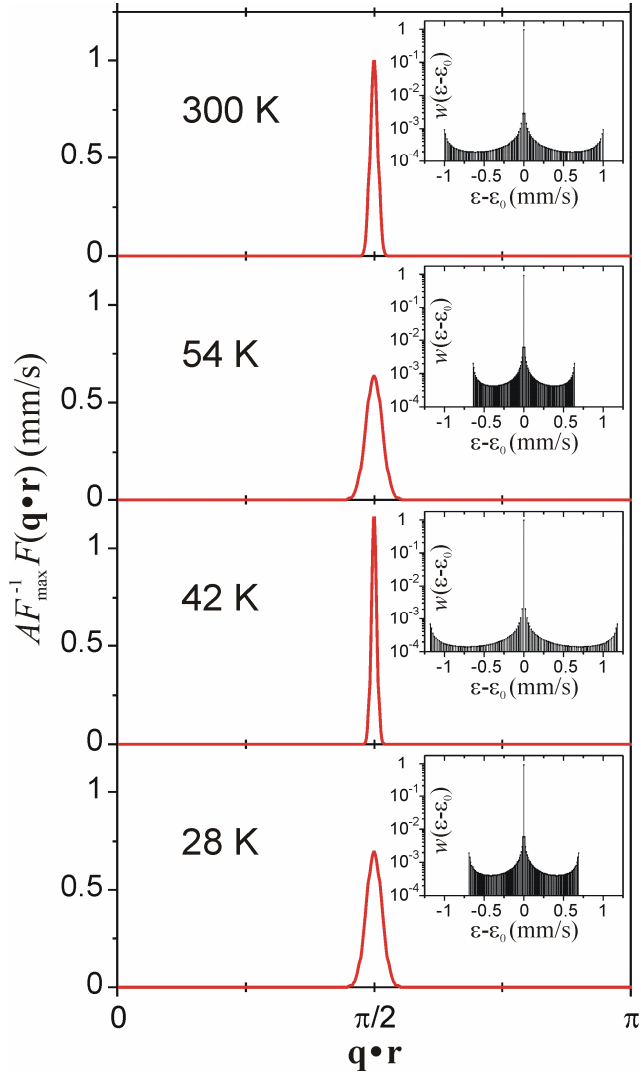

**Figure 6** Shape of the EFGW $AF_{max}^{-1}F(\mathbf{q}\bullet\mathbf{r})$ versus phase angle $\mathbf{q}\bullet\mathbf{r}$ for various temperatures. Corresponding insets show distributions $w(\varepsilon-\varepsilon_0)$ of the quadrupole shift $\varepsilon-\varepsilon_0$ with the assumption that $\varepsilon_0=0$.